# Classification of Informative Frames in Colonoscopy Videos Using Convolutional Neural Networks with Binarized Weights

Mojtaba Akbari, Majid Mohrekesh, Shima Rafiei, S.M. Reza Soroushmehr, Nader Karimi,
Shadrokh Samavi, Kayvan Najarian

*Abstract—* Colorectal cancer is one of the common cancers in the United States. Polyp is one of the main causes of the colonic cancer and early detection of polyps will increase chance of cancer treatments. In this paper, we propose a novel classification of informative frames based on a convolutional neural network with binarized weights. The proposed CNN is trained with colonoscopy frames along with the labels of the frames as input data. We also used binarized weights and kernels to reduce the size of CNN and make it suitable for implementation in medical hardware. We evaluate our proposed method using Asu Mayo Test clinic database, which contains colonoscopy videos of different patients. Our proposed method reaches a dice score of 71.20% and accuracy of more than 90% using the mentioned dataset.

## I. Introduction

Colorectal cancer is the third cause of death in United States in 2017 [1]. Detection of cancer at early stage of disease will increase chance of treatment and also survival of patients. Colonoscopy is preferred method for analyzing inside of human gastrointestinal tract and colorectal cancer detection. Colonoscopy suffers from human mistakes and also lack of sensitivity of operators [2]. Another method for analyzing of human gastrointestinal tract is using endoscopy capsule, which moves inside human body and sends images with wireless transmitter to the receiver for physician analysis. In both analyzing methods, resulting videos has many frames, which are not informative and will take much time for physician to analyze them. Hence, computer-aided methods, which classify each frame into informative and non-informative category, will help physician. Detection of polyps in both cases is a challenging task due to variation of polyps in shape and intensity in different frames. Hence, different methods have been proposed to solve this problem.

Two methods proposed in [3] for classification of informative frames. First method works based on edges and second method uses clustering of features for polyp detection. In edge-based method, they used isolated pixels in the output of canny edge detector for classification of informative frames. It introduces Isolated Pixel Ratio (IPR) as a criterion in its classification and then uses upper and lower thresholds for decision making based on the IPR criterion. Work of [3] uses block-based voting method for increasing accuracy of classification in ambiguous results of thresholding. Clustering-based method proposed by [3] uses Discrete Fourier Transform (DFT) and co-occurrence matrix for extracting features and a two-stage K-means clustering for classification of frames. In the first stage of clustering, each frame will be classified into informative, non-informative and ambiguous classes and second stage of clustering will classify each ambiguous frame into informative and non-informative classes.

Another method in detection of informative frames is Feature extraction and using trainable classifier. The method proposed in [4] uses 2D DCT transform for extraction of features from each image patch. Patch extraction proposed by [4], extracts non-overlapping patches from each image and uses dominant DCT coefficients for reconstructing each patch. Putting all reconstructed patches together will produce reconstructed image. Feature extraction method in [4] calculates difference between input image and reconstructed image in multi scales and uses their histogram as a feature. Another proposed feature in [4] is energy of all 3×3 patches in difference image. Finally, classification is done by using random forest classifier. The method proposed in [5] uses two cascaded SVM classifiers for classification of output frames of wireless capsule endoscopy. In both stages of cascaded classifiers, [5] uses non-linear SVM, first stage trained by features in histogram of HSV color space and second stage trained by multiresolution features. Feature extraction method proposed in [6] uses wavelet transform. It also uses Bayesian classifier for frame classification. Wavelet transform will decompose image into two regions. First region contains low frequency information called approximation sublayer and second region contains high frequency information called detailed sublayer. [6] converts input image into HSV color space and uses Haar wavelet transform in each color space channel and then uses $L^2$ norm of detailed sublayer of each channel as a feature for classification. Feature extraction proposed in [7] uses detailed sublayer of wavelet transformed image and co-occurrence matrix and then uses LDA classifier for classification of frames. Proposed feature extraction method in [8] is based on geometrical and textural features.

Convolutional neural network (CNN) is a type of deep learning method recently used for machine vision applications and outperforms most of pervious methods proposed for image processing tasks like semantic

Mojtaba Akbari, Majid Mohrekesh, Shima Rafiei, and Nader Karimi, are with the Department of Electrical and Computer Engineering, Isfahan University of Technology, Isfahan 84156-83111, Iran.

S.M. Reza Soroushmehr is with the Department of Computational Medicine and Bioinformatics and Michigan Center for Integrative Research in Critical Care, University of Michigan, Ann Arbor, U.S.A.

Shadrokh Samavi is with the Department of Electrical and Computer Engineering, Isfahan University of Technology, Isfahan 84156-83111, Iran. He is also with the Department of Emergency Medicine, University of Michigan, Ann Arbor, U.S.A.

Kayvan Najarian is with the Department of Computational Medicine and Bioinformatics, Department of Emergency Medicine and the Michigan Center for Integrative Research in Critical Care, University of Michigan, Ann Arbor, U.S.A.

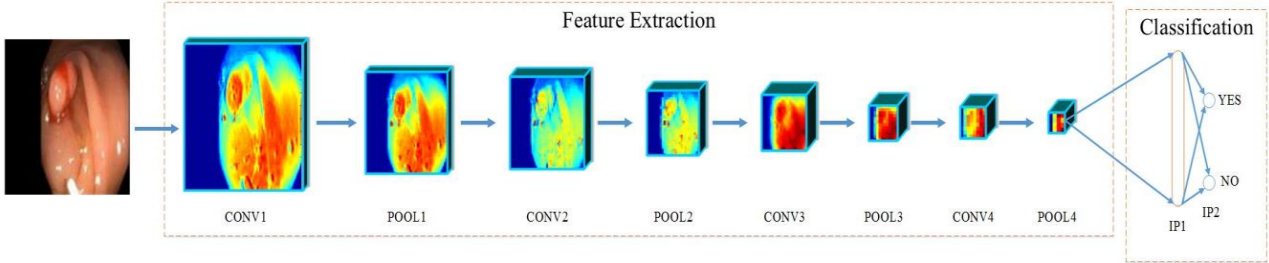

Figure 1. Structure of the proposed CNN for classification of colonoscopy frames

Table I. Parameters of the proposed CNN structure

| Layer | Parameters | | | |
|---|---|---|---|---|
| | Type | Size | Stride | Features |
| **CONV1** | Convolution | 3×3 | 1 | 8 |
| **POOL1** | Max-Pooling | 2×2 | 2 | - |
| **CONV2** | Convolution | 3×3 | 1 | 16 |
| **POOL2** | Max-Pooling | 2×2 | 2 | - |
| **CONV3** | Convolution | 5×5 | 1 | 32 |
| **POOL3** | Max-Pooling | 2×2 | 2 | - |
| **CONV4** | Convolution | 5×5 | 1 | 32 |
| **POOL4** | Max-Pooling | 2×2 | 2 | - |
| **IP1** | Fully Connected | - | - | 128 |
| **IP2** | Fully Connected | - | - | 2 |

## II. PROPOSED METHOD

Our proposed polyp detection method is based on classifier CNN that predicts label of input image based on trained features. We also binarized our proposed CNN structure for implementing on available medical hardware.

### A. Proposed CNN Network

Our proposed CNN structures contains four convolutions and pooling layers. In most CNN structures convolution layer comes with pooling layer to reduce size of data for further analysis in other layers. Convolution layer contains different filter banks and learns different features based on input patches. Pooling layer always reduces data size based on stride and method of pooling. CNNs can be used as a classifier for labelling input images based on extracted features like LeNet that uses for handwriting detection on MNIST database. Later version of CNN also contains deconvolution layers to predict and segment input images based on their ground truth like FCN [16] and U-Net [17]. Our proposed CNN architecture belongs to group of classifier CNNs and we do not have any deconvolution layer. We used two fully connected layers after convolution and pooling layers for classification of extracted features.

Fig. 1 shows block diagram of proposed CNN architecture for polyp detection in colonoscopy frames and Table I shows parameters of each layer in proposed CNN architecture.

### B. Binarization

In recent years many algorithms for reduction of memory consumption were applied in the training phase of CNNs. In this paper we use the method of [14] for binarization which results approximately 32 times smaller network. Then by using subtraction operations instead of multiplications the obtained speed is doubled. We approximate weights of each convolution layer using (1), where $W$ is weight matrix, $\alpha$ is a floating point number and $B \in \{\pm 1\}^{c \times h \times l}$ is binary approximation of the weights [14]. Also, $c$, $h$ and $l$ are the size of kernels for each convolution layer.

$$W \approx \alpha B \qquad (1)$$

Finding $\alpha$ and $B$ is an optimization problem that can be interpreted with the Equation 2 [14]. Minimizing cost function in Equation 2 with respect to $\alpha$ and $B$ results in optimum value for these parameters.

segmentation and different biomedical image processing tasks [9]. Training a CNN is also very important and challenging specially in medical applications. [10] evaluates both full training and fine tuning methods in training of CNN for medical applications like colonoscopy polyp detection. The method proposed in [11] also uses CNN with three convolution and pooling layers and fully connected layer for detection of polyp in colonoscopy images. Implementing of CNNs in hardware is also very hard because of numerous number of weights and kernels. For example AlexNet is one of the famous CNNs which needs 249MB of memory for saving weights and kernels [12]. One of the applications for binarizing the weights in medical application is the method proposed in [13] which uses weight binarization method proposed by [14] to decrease storage needed for saving weights and kernels in hand gesture recognition. Binarizing the weights and kernels would be very useful for implementing CNN on hardware.

In this paper we propose a novel polyp detection method based on CNN and also binarization of weights and kernels to decrease required memory and enables our CNN to be implemented on different hardwares. Previous methods in this field need to be improved because of lack of sensitivity and their unstable results for different structures of polyps. In section II we introduce our proposed CNN structure and also binarization method we used for binarizing the weights and kernels in training phase of CNN. We also evaluate our proposed method in section III for different quality assessment metrics on Asu Mayo Test database [2], [15]. Our proposed method outperformed the previous method in this database.

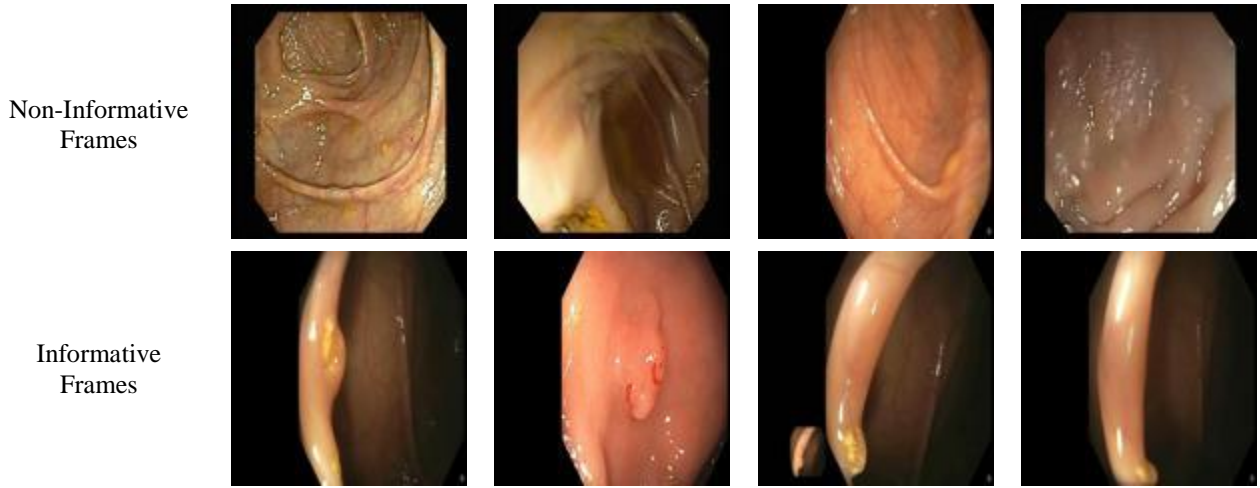

Figure 2. Examples of database images and detected labels.

$$J(B, \alpha) = \|W - \alpha B\|^2 \quad (2)$$
$$s.t.\ B \in \{\pm 1\}^{c \times h \times l}, \alpha \in R$$

Solving Equation 2 with respect to $\alpha$ and $B$ will results in following equations for them as shown in Equation 3 and Equation 4.

$$\alpha^* = \frac{1}{n}\|w\|_{l1} \quad (3)$$

$$B^* = sign(W) \quad (4)$$

Using this method in CNN will change all floating point numbers to binary numbers and reduces size of network. Thus, for each kernel we will have binary weights and a scaling factor $\alpha$.

We assumed our fully connected layer as a convolution layer with the number of features equal to number of neurons. Thus, all layers in networks will be binarized except their bias values.

We have shown in [13] that using binarized values of weights will slow down the speed of learning procedure and also increases number of iterations needed for convergence of learning. In order to outcome these problems we store floating point values of weights and update all of them in each iteration but the gradient will be calculated based on the binary values of weights in each iteration and just binarized values will be used in test phase.

## III. EXPERIMENTAL RESULTS

We used Asu Mayo Test clinic database for evaluating our proposed method in both polyp detection and binarization [15]. This database contains 18 colonoscopy videos which nine of them contain polyps and others are normal. All has ground truth data for polyp segmentation annotated by expert physicians. Resolution of videos is different and database contains both high and low resolution videos. We extracted all frames in videos for evaluating our proposed method and used 14 videos for train and 4 videos for test phase. In both phases, half of videos contain polyps and the other half are normal. Then we shuffled frames of training videos randomly selected them for time priority in training and then tested our algorithm with the trained FCN. Our training database contains 12872 frames which 3486 frames of them contain polyps and testing database contains 4702 frames which 827 frames of them has polyps inside of the frame.

We trained our proposed CNN architecture and performed the binarization using Caffe [18] library and MATLAB R2014a interpreter. Because of variation in size of frames of different videos, we resize them to 118×118 images and feed them to the network with their corresponding labels. Labels of frames are based on their corresponding ground truth data. Fig. 2 shows some examples of colonoscopy images in the database and their labels which are correctly detected.

Binarizing of CNN weights will effect on performance in test phase. Table II demonstrates the evaluation of our proposed method in both binarized and floating point weights. We fixed 90,000 iterations for both scenarios and saved trained weight in each 10,000 iterations. Data of Table II is average values for resulting accuracy, recall and False Positive Per Frames (FPPF) criteria in both scenarios. Table II proves that binarizing weights in CNN will reduce performance in terms of mentioned criteria, but the difference for both scenarios is negligible and guarantees that our proposed method can be used in hardware with binarized weight.

For the sake of comparison between the performance of our proposed method and the polyp detection method, proposed in [11], we implemented this method as well. This algorithm is a CNN based method and also designed for detecting polyps in colonoscopy videos and uses three convolutions and pooling layers for feature extraction and fully connected layer with 256 nodes followed by another fully connected layer with two nodes for classification. We trained both CNN architectures with same training sets in 2,000 iterations. We saved weights and tested both methods with the same testing set. Table III shows results of both methods for different metrics.

Table II. Comparison Between Floating Weights and Binarized Weight in Polyp Detection

| Weight | Accuracy (%) | Recall (%) | FPPF |
|---|---|---|---|
| Floating Point | 90.4 | 69.3 | 0.04 |
| Binarized | 90.0 | 79.2 | 0.06 |

Table III. Comparison of the proposed method with method of [11]

| Criterion | Accuracy (%) | Dice Score (%) | Recall (%) | Precision (%) | Specificity (%) | FPPF |
|---|---|---|---|---|---|---|
| *Proposed Method with Binarization* | 90.28 | 71.20 | 68.32 | 74.34 | 94.97 | 0.06 |
| *Shin et al [11]* | 86.69 | 43.30 | 28.90 | 86.28 | 99.02 | 0.05 |

IV. CONCLUSION

In this paper we proposed a novel polyp detection method based on CNN. We used four convolutions and pooling layers followed by two fully connected layers in feature extraction and classification of proposed architecture respectively. We also binarized the weights and kernels to reduce the size of our proposed network. Binarizing the weights and kernels also makes our proposed network ready for implementing on medical hardware. We also numerically evaluated our proposed method in both floating point and binary weights on Asu Mayo Test clinic database. Our proposed method reached appropriate numerical results and also outperformed previous method of polyp detection in colonoscopy images.

V. FUTURE WORK

Binarization of weights and kernels will reduce size of networks and also it would be the first step in implementation of network inside portable medical hardware. Therefore, our proposed method can be used in colonoscopy devices and in future endoscopy capsules for transmitting informative frames to a receiver. Using our method will also decrease power consumption in endoscopy capsules which is an important factor in the design of these devices. Our next step would be the hardware implementation of the proposed method.